\documentclass[epj]{webofc}
\usepackage[varg]{txfonts}   
\usepackage{graphicx}
\wocname{EPJ Web of Conferences}
\woctitle{ICNFP 2015}
\begin{document}
\selectlanguage{english}
\title{Confining potential of Y-string on the lattice at finite T}
%
%

\author{Ahmed S. Bakry\fnsep\thanks{\email{abakry@impcas.ac.cn}}, Xurong Chen, and Peng-Ming Zhang} 

\institute{Institute of Modern Physics, Chinese Academy of Sciences, Lanzhou 730000, and State Key Laboratory of Theoretical Physics, Institute of Theoretical Physics, Beijing 100190, China.}

\abstract{The potential due to a system of three static quark $(3Q)$ is studied using SU(3) lattice QCD at finite temperature with Polyakov loops operators. We focused our analysis on the large distance properties of the $3Q$ potential and found a good fit behavior to the Y-string model formula. In addition to the linearly confining term proportional to the minimal length of the Y-string, we observed that the subleading logarithmic term, which is proportional to Dedekind $\eta$ function and accounts for the Y-string's quantum fluctuations, is necessary to reproduce the quark anti-quark $Q\bar{Q}$ string tension of the corresponding mesonic system at finite temperature.}
\maketitle
\section{Introduction}
\label{intro}
  The three-quark confining potential is a fundamental quantity in elementary particle physics and is essential for understanding baryon structure~\cite{LlanesEstrada:2011kc}. It remains, however, an insurmountable task to put an analytic form to the binding forces between quarks in the non-perturbative region of QCD from first principles. 

  Perturbative QCD provides a good description to the short-distance aspects of the $3Q$ potential as a two-body (Coulombic) one-gluon-exchange (OGE) interaction potential. More recently, it has been shown~\cite{Brambilla:2009cd} that the breakdown of the two body force in the short-range happens at two loops when the first genuine three-body force appears. Discussions concerning the intermediate and large distances, however, are usually carried out either on phenomenological bases~\cite{Andreev:2015riv,Takahashi:2002bw,deForcrand:2005vv,Jahn2004700}, making use of the strong-coupling expansion~\cite{PhysRevD.11.395,Nakano:2009bf} arguments or with the lattice calculations of the three-quark Wilson loop~\cite{Takahashi:2000te}. 
 
  Lattice simulations have shown that the $Q\bar{Q}$ potential is linearly rising~\cite{Creutz:1980zw}. In a string phenomenology, the linear rise is consistent with the formation of a stringlike flux tube linking the color sources. The quantum fluctuations of the string produce the well-known Coulomb-like sub-leading corrections to the $Q\bar{Q}$ potential, namely, the L\"uscher term. Simulations of many confining gauge models~\cite{luscher,Juge:2002br,Bali,Pennanen:1997qm,Caselle:1995fh,Gliozzi:2010zv,Gliozzi:2010jh} have verified the existence of this term.
                     
 In the baryon, the fit analysis of the lattice data would suggest two types of parametrization~\cite{Takahashi:2000te} depending on the interquark distances,~i.e., the $\Delta$ and Y-type potential. The $\Delta$-potential describes a sum of two-body forces and is proportional to the perimeter of the $3Q$ triangle with a string tension half that of the corresponding $Q\bar{Q}$ system. On the other hand, the Y-potential approximates a three-body force proportional to the minimal length of the three strings with the same string tension as the corresponding $Q\bar{Q}$ system. In this picture, the three body force is conceived as a result of the formation of a Y-shaped string linking the three-static quarks. This realization is consistent with the strong coupling approximation and with the dual superconducting picture of QCD~\cite{PhysRevD.11.395,PhysRevD.34.2809,Brambilla1995113}.

  However, many subtleties in the lattice simulations have made the exact distances at which each behavior takes over a subject of debate~\cite{sommer,Thacker,Alexandrou:2002sn,Takahashi:2000te,Takahashi:2002bw, Andreev:2015riv}. In addition, unlike the simpler $Q\bar{Q}$ case, the sub-leading corrections arising from the quantum fluctuations of the Y-string linking the $3Q$ quark system yet has not been verified in SU(3) gauge theory. These stringy fluctuations are expected to be more enhanced at temperatures very close to the deconfinement point.   
 
  Usually in lattice simulations the potential energy is extracted from trial states that maximally overlap with the ground state using Wilson loop operators~\cite{Alexandrou:2002sn,Alexandrou:2001ip,Takahashi:2000te,Bakry:2011cn}. Otherwise, the potential may be extracted with the use of Polyakov loop operators. These thermal loops address the free-energy of the system ~\cite{Luscherfr} and are suitable operators to study lattice gauge theories coupled to heatbath.  

  At zero temperature, the numerical evaluation of the $3Q$ potential with the use of Polyakov loop operators in 4D SU(3) Yang-Mills theory is known to be computationally challenging. At finite temperature, on the other hand, the loop operators assume a short extent in the time direction. By the employment of adequate levels of UV filtering~\cite{Bakry:2010sp} such that the physical measurements of the $3Q$ potential are intact, the evaluation of the two-point correlation function to extract the potential and even the three-point function corresponding to the energy density can be viable~\cite{Bakry:2015csa,Bakry:2014gea} at finite temperature. The simulations for the energy density in the $3Q$ system, for example, have revealed a filled $\Delta$-shape profile even at large distances. 

  A more profound analysis of these filled-$\Delta$-shaped profiles revealed an underlying system of three-convoluted Y-shaped-Gaussian~\cite{Bakry:2014gea}. The energy density profile is peculiar~\cite{Bakry:2014gea} and has not been observed at zero temperature with the employment of Wilson's loop operator~\cite{Okiharu:2003vt,Bissey:2005sk,Bissey}. Nevertheless, the characteristics of the associated potential of the static $3Q$ system at finite temperature is not yet fully understood. 

  In this study, we discuss the $3Q$ potential derived from the Y-string model and investigate concisely how it compares to the corresponding lattice data of the SU(3) pure Yang-Mills theory at finite temperature. This may delineate the fine properties of the energy density and potential of the $3Q$ system even at low and zero temperatures. Our goal is to find out if the Y-law derived from the effective bosonic string model~\cite{Jahn2004700} provides a satisfactory description of the lattice data and reproduces the corresponding $Q\bar{Q}$ string tension. To scrutinize further the nature of the $3Q$ potential data, we draw a comparison with other fit ansatzes such as the $\Delta$-type potential and the so-called bare Y-ansatz with the string's sub-leading corrections suppressed.
    
  The contents are organized as follows: In Section (II), we discuss the Y-string model at finite temperatures. In section (III) we describe the $3Q$ potential measurements. The fit analysis of the lattice data is depicted in section (IV). The conclusion is provided in the last section.
 
\section{String's phenomenology of the $3Q$ potential}
 A pure Yang-Mills vacuum can admit stable string-like objects which confine static color charges and give rise to the linearly rising potential. In a mesonic string, the quantum fluctuations entail the L\"uscher subleading correction~\cite{Luscherfr,Luscher:2002qv} to the linearly rising potential. This has been found in consistency with very precise lattice measurements of the the $Q\bar{Q}$ potential for color source separation commencing from distance $R=0.4$ fm~\cite{Luscherfr}. 

 At high temperature, the thermal behavior of the free-string manifests at source separation distance scales larger than what one would expect normally in the zero temperature regime $R \approx 0.8$ fm~\cite{Bakry:2010sp}. The inclusion of higher-order string's self-interactions improves the match between the lattice data and the string model~\cite{Caselle2003499}.    

  In a static baryon, the Y-string system is expected~\cite{Jahn2004700} to be a stable configuration in the IR region of the theory. This system accounts for three strings originating from a node to the static three quarks. A theoretical development discussing the effects of the quantum fluctuations of the Y-string on the $3Q$ potential has been reported~\cite{Jahn2004700}. The calculations of the Casimir energy in this model have indicated a geometrical L\"uscher-like sub-leading term for the Y-$(3Q)$ potential~\cite{deForcrand:2005vv}. The L\"uscher-like term is a gauge-group-independent and depends only on the geometry of the three static charges. 
    
  The simulation of 3-Potts model showed a $3Q$ potential~\cite{deForcrand:2005vv} consistent with the L\"uscher-like correction $\gamma/L_{Y}$ to the linearly rising Y-ansatz, and hence the applicability of the Y-baryonic string picture at large distances. 

  The Y-string may be described by the Nambu-Goto type string action with Dirichlet boundary condition at the static color sources. In this formalism, the action is proportional to the total area of the three blade world sheet system swept by the fluctuating world lines of the strings, in addition to a boundary term which accounts for the junction fluctuations. This Y-string system~\cite{Jahn2004700} has a mixed Dirichlet-Neumann condition~\cite{tHooft} and the string's partition function $Z_i(\phi)$ in $D$ dimensions~\cite{Jahn2004700} is given by

\begin{equation}
  Z_i(\phi)=e^{-\frac{\sigma}{2}\int |\partial
\xi_{\min,i}|^2}
 |\det(-\triangle_{\Theta_i})|^{-(D-2)/2}\:,
\label{Laplace}
\end{equation}
   where $\xi_{\min,i}$ is the minimal-area solution for a given junction configuration $\phi(t)$, and $\triangle_{\Theta_i}$ denotes the Laplacian acting on the domain (blade) $\Theta_i$.

 Jahn and De~Forcrand~\cite{Jahn2004700} calculated the Casimir energy for the baryonic potential ${\rm V}_{3Q}$. This was done by evaluating the determinant of the Laplacian in Eq.~(\ref{Laplace}) by conformally mapping the resulting domains to rectangles. This results in an expression for the determinant in terms of the Dedekind $\eta$ function and a Gaussian function of the junction fluctuations, ${\bf e_i} \cdot \bf{\phi_{w_{n}}}$, which in 4D is given by 
\begin{eqnarray}
\label{barypotgaussian}
\det(-\triangle_{\Theta_i}) = \eta^2\!\left( \frac{iL_{T}}{2L_i}\right) \nonumber  \exp\left(- \frac{1}{12\pi}\sum_{w_{n}} w_n^3 \coth (w_n L_i)| {\bf {e_i}} \cdot {\bf{\phi_{w_{n}}}}|^2\right)\:,
\end{eqnarray}
  with $w_n=2\pi n/L_T$, where $L_T=1/T$ is the time extent of the blade, and $L_i$ are the lengths of the strings. The sum over all eigen-energies would result in a L\"uscher-like correction to the $V_{3Q}$ potential at zero temperature~\cite{Jahn2004700}. The baryonic potential ${\rm V}_{3Q}$ then reads

\begin{align}
\label{vba1}
V_{3Q}(L_i)=& \sigma L_Y + V_{\parallel} +2 V_{\perp} + O(L^{-2}_i)\:,\notag\\
\intertext{with}
V_{\parallel}(L_i) =& -\sum_i \frac{1}{2L_T}\eta\left(\frac{i L_T}{2L_{i}}\right)+ \sum_{w=0} \frac{1}{L_T}\ln\left[\frac 1 3 \sum_{i<j} \coth(w L_i)\coth(w L_j)\right]\:, \notag \\
\intertext{for the in-plane component and}
V_{\perp}(L_i) =& -\sum_i \frac{1}{2L_T}\eta\left(\frac{i L_T}{2L_{i}}\right)+ \sum_{w=0} \frac{1}{L_T} \ln\left[ \frac 1 3 \sum_i \coth(w L_i)\right]\:,
\end{align}

\noindent is the potential component due to the perpendicular fluctuations such that $L_{\rm Y} \equiv L_1+L_2+L_3$ is the minimal of the sum of the lengths of the three strings. The corresponding mesonic limit would read   

\begin{align}
V_{\perp}= -\frac{1}{2L_T}&\eta \left(\frac{iL_T}{2L_{1}}\right)- \frac{1}{2L_T}\eta \left(\dfrac{i L_T}{2 L_{2}}\right)+\sum_{w=0} \frac{1}{L_T} \ln\left[ \frac 1 2 \coth(w L_1)+\coth(w L_2) \right].\:
\end{align}
 The quark anti-quark ($Q\bar{Q}$) potential is then
\begin{equation}
V_{Q\bar{Q}}=\sigma (L_1+L_2)+ \frac{1}{L_T}\ln\left[\eta \left(\dfrac{i L_T}{2(L_{1}+L_{2})}\right)\right],
\label{meson}
\end{equation}

\noindent which is in agreement with the mesonic string potential~\cite{Caselle2003499}. Expressing the sum in Eq.~\eqref{vba1} in terms of Dedekind $\eta$ functions, the potential in the $3Q$ channel would read

\begin{equation}
V_{3Q}=\sigma L_{\rm Y}- \dfrac{\gamma}{L_T} \ln\left[\eta \left(\dfrac{i L_T}{2L_{\rm Y}}\right)\right],
\label{total}
\end{equation}

\noindent where $\gamma$ is a geometrical factor that can be evaluated numerically by solving Eq.~\eqref{vba1}. 

\section{The lattice QCD measurement for the 3Q potential} 
  A transfer matrix interpretation to the Polyakov loops correlator allow to obtain the $3Q$ static potential $V_{\rm 3Q}$ with the center symmetry preserving operator 

\begin{align}
\langle \mathcal{P}_{3Q}\rangle &= \langle \mathcal{P}(\vec{r}_{1} ) \mathcal{P}(\vec{r}_{2} ) \mathcal{P}(\vec{r}_{3} ) \rangle, \nonumber\\
 &= \exp(-V_{3Q}(\vec{r}_{1},\vec{r}_{2},\vec{r}_{3};T)),
\label{c}
\end{align}

\noindent where the Polyakov loop is given by

\begin{equation}
   P(\vec{r}) = \frac{1}{3}\mbox{Tr} \left[ \prod^{N_{t}}_{n_{t=1}}U_{\mu=4}(\vec{r},n_{t}) \right].
\end{equation} 

  In the context of the Polyakov loop method we have contributions of temperature-dependent effects to the free energy of a system of three static charges coupled to a heatbath. In the following we describe the lattice arrangement and parameters used to extract the above correlator Eq.~\eqref{c} at finite temperature.


  The gauge configurations were generated using the standard Wilson gauge action. The two lattices employed in this investigation are of a typical spatial size of $3.6^{3}\rm{fm}^{3}$. Performing the simulations on large enough lattice sizes would be beneficial to gain high statistics in a gauge-independent manner and also minimizing the mirror effects and correlations across the boundaries as a by-product~\cite{Bali:1994de,Bissey}.

  The SU(3) gluonic gauge configurations have been generated employing a pseudo-heatbath algorithm~\cite{Fabricius,Kennedy} updating the corresponding three SU(2) subgroup elements \cite{Cabibbo}. Each update step consists of one heatbath and 5 micro-canonical reflections. We chose to  perform our analysis with lattices as fine as $a = 0.1$ fm by adopting a coupling of value $\beta = 6.00$, with temporal extents of $N_{t}=8$, and $N_{t}= 10$ slices, which correspond to temperatures $T/T_{c} \simeq 0.9$, and $T/T_{c} \simeq 0.8$, respectively. 

  We perform a set of measurements $n_{\mathrm{sub}}=20$ separated by 70 sweeps of updates. Each set of measurements is taken following  2000 updating sweeps. These sub-measurements are binned together in evaluating Eq.~\eqref{c}. The total measurements taken on 500 bins. In this investigation, we have taken 10,000 measurements at each temperature. The measurements are taken on hierarchically generated configurations.

  An ultraviolet (UV) filtering step~\cite{PhysRevD.82.094503} precedes our measurements of the three Polyakov loop correlator Eq.~\eqref{c}. We have previously shown that with adequate levels of 4D smearing the effects on the $3Q$ potential can be kept intact. We have shown that this is consistent with a mean square diameter of the Brownian motion smaller than any interquark separation. 

\section{Numerical results and discussions}
  For the study of the $3Q$ potential $V_{3Q}$, we are interested in the large-distance behavior which is directly related to the properties of the confining force rather than the short-distance behavior~\cite{Takahashi:2000te} that is well described by the two-body Coulomb-type potential.

\begin{figure}[!hpt]
\begin{center}
\includegraphics[width=6.5cm,height=5.0cm]{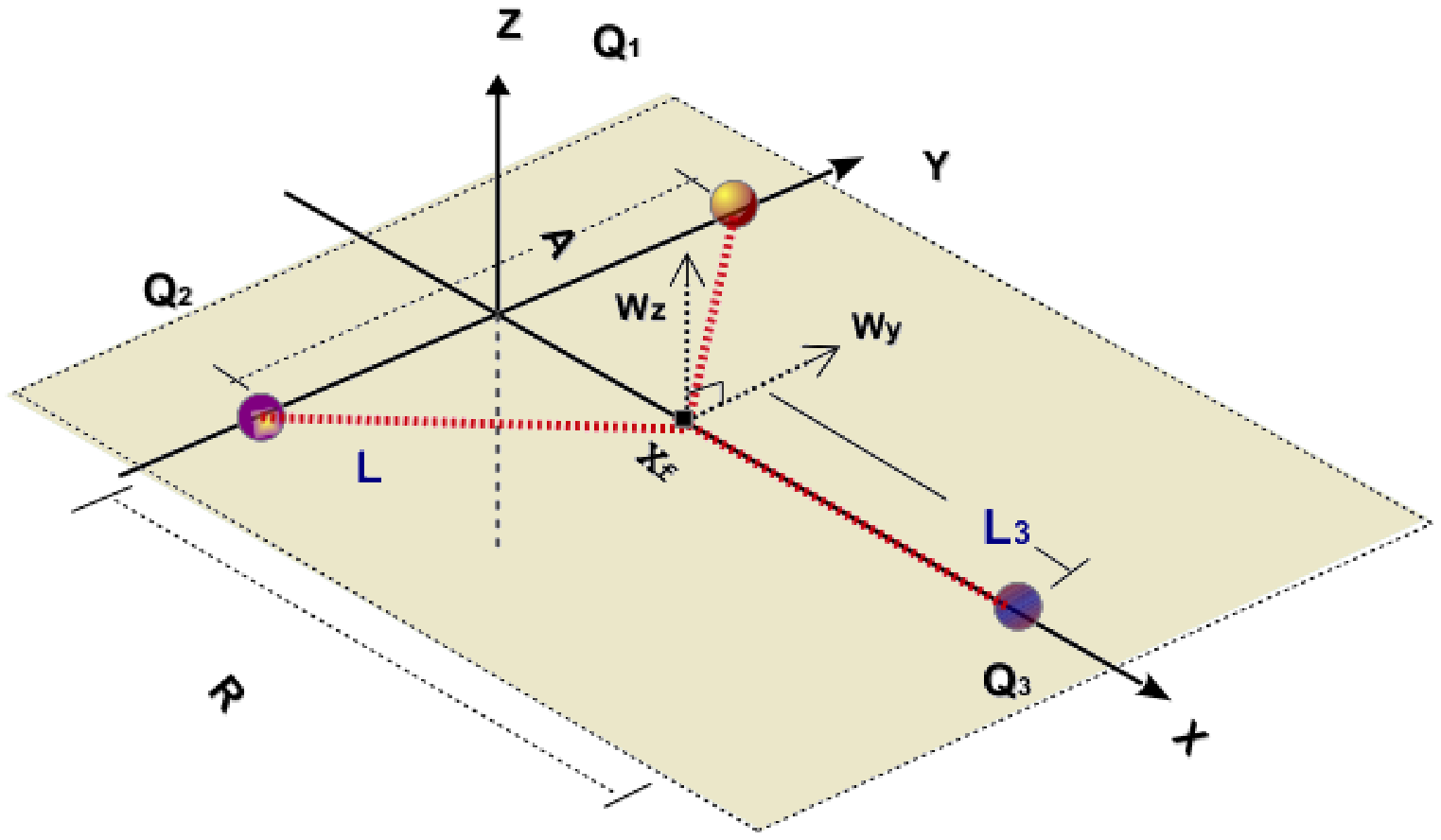}
\includegraphics[width=6.0cm,height=5.0 cm] {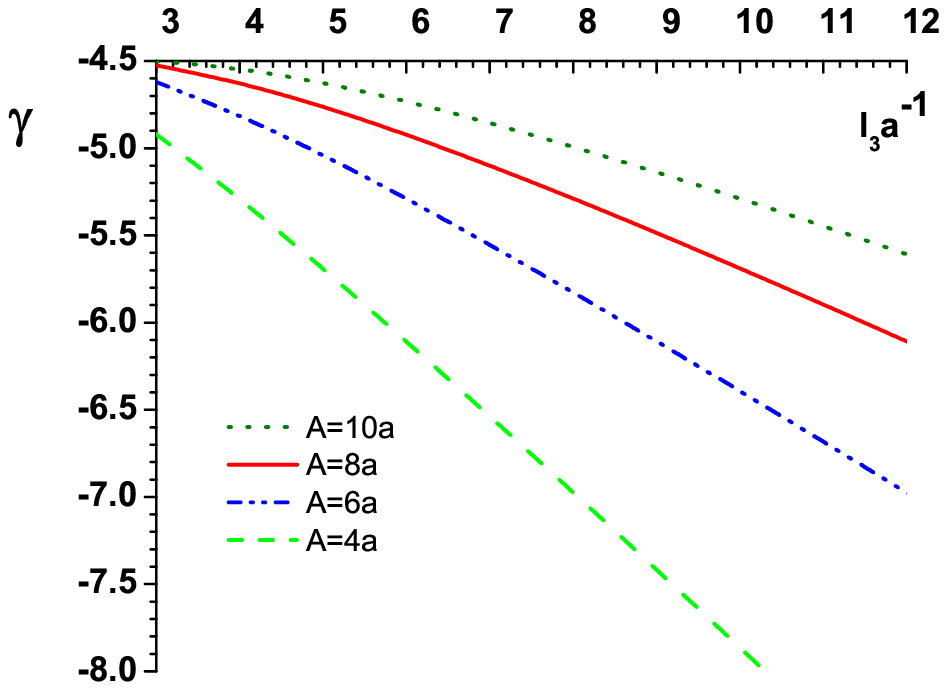}
\caption{(Left) Schematic diagram showing the configuration of the $Y$ string relative to the quark source positions. The junction’s locus is fixed at Fermat point $x_f$. (Right) The geometrical factor $\gamma$ of the log term in Eq.~\eqref{total}. The solid lines correspond to $\gamma$ as a function of the third string length $L_3$
for three isosceles triangles with bases of length $A=0.6$,  $A=0.8$ fm and $A=1.0$ fm. The solid lines correspond to $\gamma$ as a function of the third string length $L_3 a^{-1}$.}
\label{gam}
\end{center}
\end{figure}

\begin{table*}[!hpt]
\caption{\label{T08}Lists the returned values of the $\chi_{\mathrm{dof}}^{2}(x)$ for fits of the lattice data to the string model formula Eq.~\eqref{total}, the fits are for isosceles triangle quark configurations of base width $A=6\,a$ and $A=8\,a$ at two temperatures.}
\begin{center}
\begin{tabular}{|c   c   c   c   c   c   c   c  c  c  |c    c  c  c}
\hline
Fit Range                     &               R=5-12&       &   &        R=6-12&       &   &        R=7-12&    &   \\\hline
Fit Parameters    & $\chi^{2}$ & C &    &$\chi^{2}$& C &  & $\chi^{2}$ & C &\\\hline
$T/T_c=0.8$                   &        &       &       &     & && & & \\
$A=6a$                              & 15.06 & -4.17   &&  7.61 &-4.17  &&   1.93  & -4.16 &   \\
$A=8a$                              & 3.99 & -3.82    &&  2.1  &-3.83  &&   1.38  & -3.84&   \\\hline
$T/T_c=0.9$                   &        &       &       &     & && & &    \\
$A=6a$                        &     46.0  & -3.05 &&     18.6  &  -3.04 &&    5.50  & -3.03 &   \\
$A=8a$                        &     2.3   & -2.77 &&     2.27  & -2.77  &&    2.26  & -2.77 &   \\\hline
\end{tabular}
\end{center}
\end{table*}
  
   We evaluate the correlator Eq.~\eqref{c} to extract the heavy quarks potential for a planar $3Q$ arrangement corresponding to isosceles triangles of base width $A$ and height $R$ as illustrated in Fig.~\ref{gam}. Figure~\ref{gam} also shows the numerical values of the geometrical factor $\gamma$ as a function of triangle length $L_3$ for isosceles configurations $L_1=L_2$ evaluated by solving Eq.~\eqref{vba1} for the Y-string potential.   
  
\begin{figure}[!hpt]
\includegraphics[width=7.0cm] {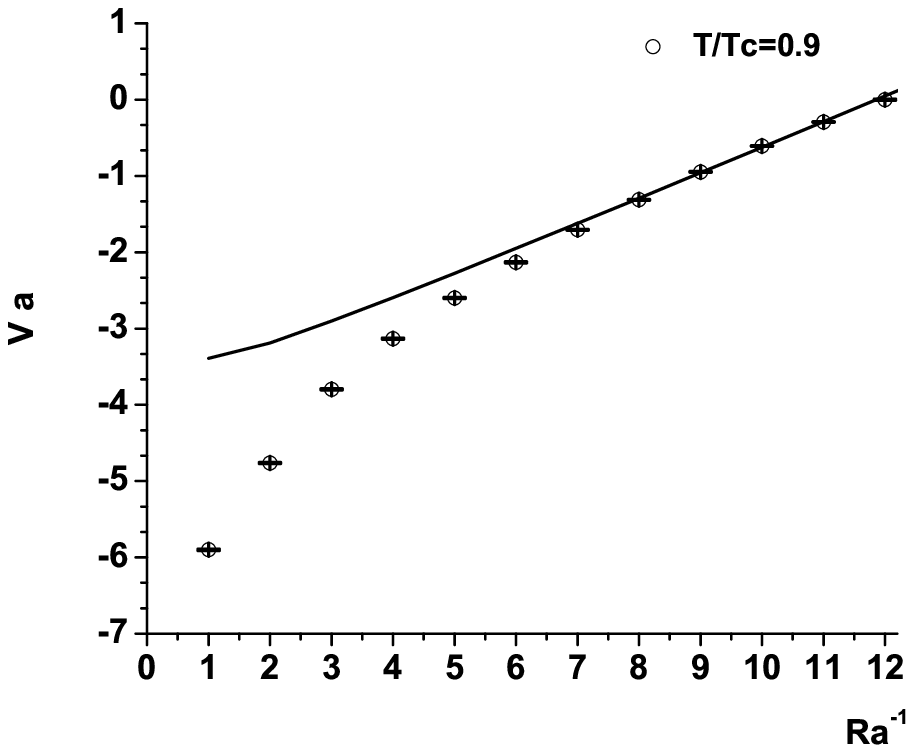}
\includegraphics[width=7.5cm] {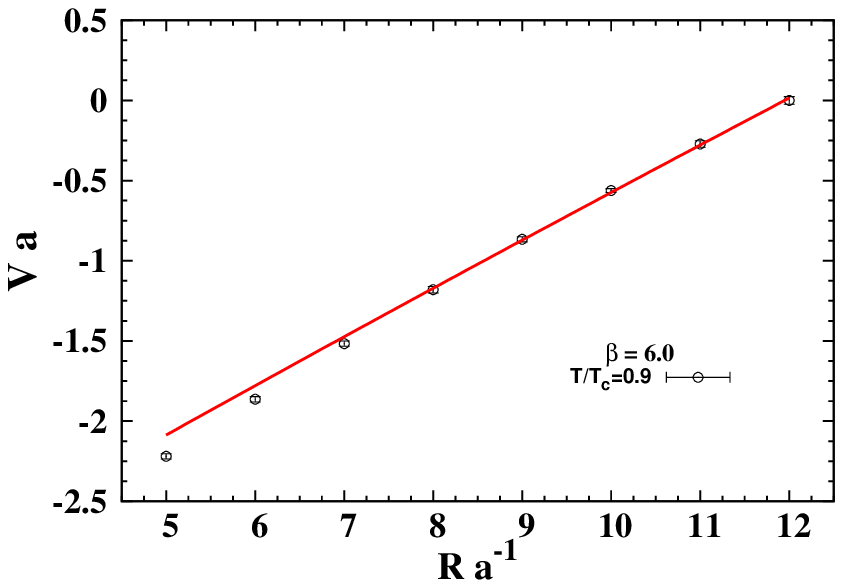}
\caption{(Left) The lattice data of the $Q\bar{Q}$ potential at $T/T_c=0.9$. (Right) the corresponding $3Q$ potential of a planar $3Q$ isosceles triangle with a base length $A=0.6$ fm. The solid lines correspond to the best fits to the string model Eq.~\eqref{meson} and Eq.~\eqref{total}, respectively.}
\label{p1}
\end{figure}

\begin{figure}[!hpt]
\includegraphics[width=7.5cm] {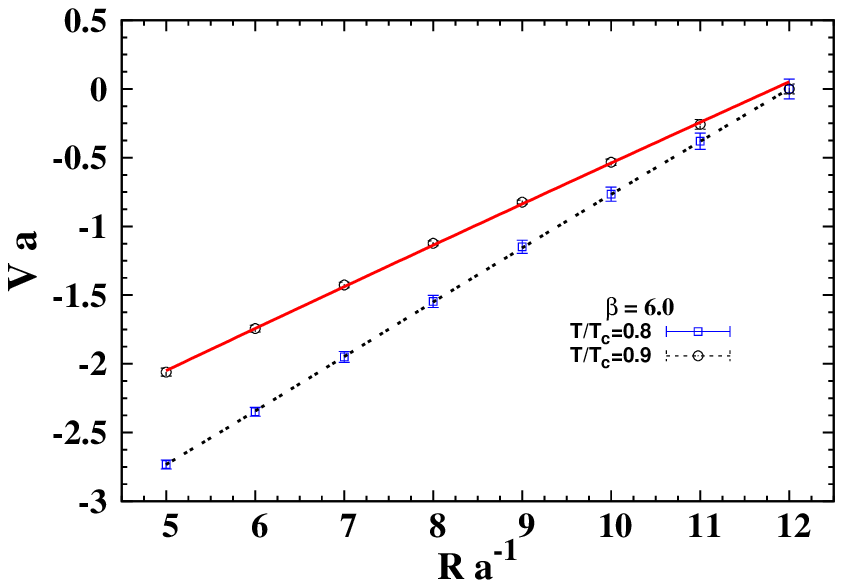}
\includegraphics[width=7.5cm] {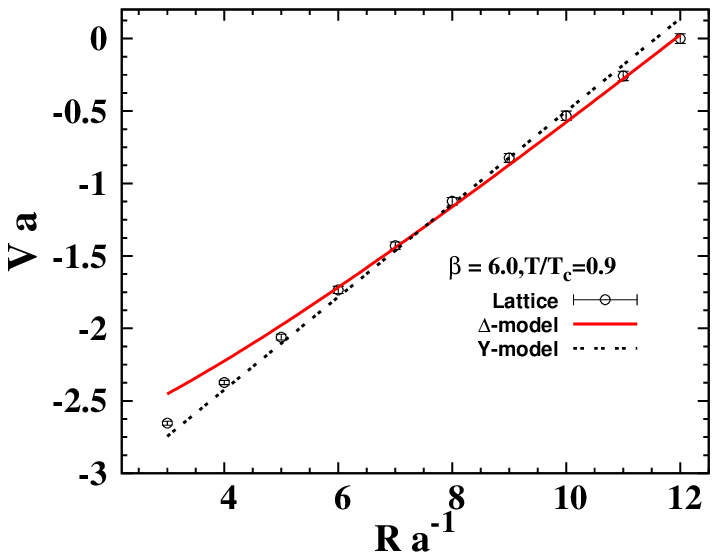}
\caption{The $3Q$ potential versus the height length of isosceles triangular configuration of the planar $3Q$ with bases of length $A=0.8$ fm. (Left) The lines correspond to the $3Q$ potential according to the best fits to the string picture formula of Eq.~\eqref{total}. (Right) The lines correspond to fits to Eq.~\eqref{d} and Eq.~\eqref{y} for fit range $R\in[5,12]$.} 
\label{p2}
\end{figure}

  The measured numerical values of the potential are reported in Fig.~\ref{p1} and Fig.~\ref{p2} for isosceles $3Q$ triangles $3Q$ corresponding to bases of width, $A=0.6\, \rm{and} \,0.8$ fm, at the two considered temperatures $T/T_c=0.8$ and $T/T_c=0.9$. The numerical values of the potential have been normalized to their values at isosceles height $R=1.2$ fm.

  The potential data for each base are extracted in accord to Eq.~\eqref{total} by varying the height of the triangle, $R$, for each fixed base, as shown in Fig.~\ref{gam}. Our consideration of different isosceles $3Q$ configuration is intended to enable a systematic identification of the distance scales at which the string model holds. 

  Formula Eq.~\eqref{total} sums up the contributions to the $3Q$ potential resulting from both the in-plane and the perpendicular fluctuations as discussed above. Before fitting this string potential to the lattice data, the Y-string’s configuration has to be fixed to its minimal value, i.e., with the node's position at Fermat point of the $3Q$ triangular configuration. The position of the Fermat point of the planar isosceles arrangement is given by $R=\dfrac{A}{2 \sqrt{3}}$. 

  Table~I summarizes the returned values of $\chi^{2}$ from the resultant fits of the $3Q$ potential at both temperatures. The rescaling parameter $C$ is the only fit parameter used. 

  In general, the fits show strong dependency on the fit range with the inclusion of the points at small $Q_3$ sources separations. Figure~\ref{p1}, for example, depicts the potential data corresponding to an isosceles $3Q$ triangle of the smallest width considered $A=0.6$ fm and the corresponding fits to Eq.~\ref{total}. The points at small isosceles height $R=0.5$ fm and $R=0.6$ fm deviate from the Y-string model curve. These deviations at relatively small heights and bases can be conceived as relevant to self-interactions of the two strings connecting the two quarks at the base, $Q_1$ and $Q_2$, in addition to the interaction of the node with the third quark $Q_3$. That is, higher order effects are expected to be more pronounced at higher temperatures.

   The string length in the case of the Y-string linking any two quark sources is  greater than the mesonic string of a corresponding $Q\bar{Q}$ pair. A fit of the potential to the mesonic string potential Eq.~\eqref{meson} is shown in the right graph of Fig.\ref{p1}. The best fits are returned if only the last four points are included, i.e., from quark separation distances $R=0.8$ fm to $R=1.2$ fm. The value returned for the string tension $\sigma_{Q\bar{Q}} a^{-2}=0.032$ is the one used as input for the string tension in formula Eq.~\ref{total} for the baryonic Y-string potential.    
 
  With the increase of the width of the base of the isosceles to $A=0.8$ fm the values of $\chi^{2}$ in Table~I reduce even for small isosceles heights of $R=0.5$ fm and $R=0.6$ fm. A wider base of the isosceles triangle would increase the length of the string linking any two quarks $L_1+L_3$. This approaches the length at which the corresponding free mesonic string commence to match the lattice data. One can also add to this the observation that the self-interactions of the adjacent strings in the Y-string configuration are expected to be negligible with the increase of the source separation at the base.
 
  We list in Table~II  values of the $\chi^{2}$ returned from fits of the $3Q$ potential to a $\Delta$-ansatz ~\cite{Takahashi:2000te,Takahashi:2002bw} given by the form  
\begin{equation}
V_{\rm 3Q}(\vec{r}_{1},\vec{r}_{2},\vec{r}_{3})=-\dfrac{1}{2}A_{Q\bar{Q}} \sum_{i<j}\frac1{|\vec{r}_{i}-\vec{r}_{j}|}+\frac{1}{2}\sigma_{Q\bar{Q}} \sum_{i<j} |\vec{r}_{i}-\vec{r}_{j}|.
\label{d}
\end{equation}
  The constant $A_{Q\bar{Q}}$ determines the strength of the OGE Coulombic term derived from perturbative QCD (see Ref.~\cite{Takahashi:2000te,Takahashi:2002bw}). The value of the string tension is taken to the half that of the mesonic string. To appreciate the effects of the junction fluctuations we consider also the fits to a Y-ansatz which in addition to the perturbative Coulombic term includes only the linearly confining term corresponding to the classical configuration of the string $L_{Y_{min}}$ with the sub-leading corrections of Eq.~\eqref{total} suppressed, i.e.,
  \begin{equation}
V_{\rm 3Q}(\vec{r}_{1},\vec{r}_{2},\vec{r}_{3})=-\dfrac{1}{2}A_{ Q\bar{Q}}\sum_{i<j}\frac1{|\vec{r}_i-\vec{r}_j|}+\sigma_{\rm Q\bar{Q}} L_{\rm Y}.
\label{y}
\end{equation}
\begin{table*}[!hpt]
\begin{center}
\caption{\label{T08} The returned $\chi_{\mathrm{dof}}^{2}(x)$ for fits of the lattice data to $3Q$ isosceles of width $A=0.8$ fm at $T/T_{c}=0.9\,$. The fits compare Eq.~\eqref{d} for the $\Delta$-ansatz, Eq.~\eqref{total} for the Y-string model and Eq.~\eqref{y} for a bare Y-ansatz.}
\begin{tabular}{|c   c   c   c   c   c   c   c  c  c  |c    c  c  c}
\hline
Fit Range                     &               R=5-8&       &   &        R=8-12&       &   &        R=5-12&    &   \\\hline
Fit Parameters    & $\chi^{2}$ &  &    &$\chi^{2}$&  &  &$ \chi^{2}$ &  &\\\hline
$\Delta$-ansatz                             & 27.43 &  &&  9.63 &       &&   40.5  &  &   \\
Y-string model                       & 1.25  &  &&  6.85 &       &&   8.09  &  &   \\
Y-bare ansatz                        & 2.36  &  &&  40.84&       &&   89.71 &  &   \\ \hline
\end{tabular}
\end{center}
\end{table*}

   Let us consider the highest temperature $T/T_c=0.9$ and focus our comparison on the triangular configuration of the base width $A=0.8$ fm. The values in Table~II indicate higher $\chi^{2}$ for the $\Delta$-shape parametrization if the fit range includes all the considered points $R \in [5,12]$. For points corresponding only to larger heights of the isosceles $R \in [8,12]$, we interestingly observe that the bare Y-model, accounting merely to the classical configuration, returns high values of $\chi^{2}$ compared to the Y-string model Eq.~\ref{total}. At large distances and temperatures~\cite{Bakry:2014gea} the Y-string assumes a higher energy content and broadening profile. This can be read as well from the value of the $\chi^{2}$ corresponding to the $\Delta$-parametrization Eq.~\ref{d} if the same fit range is considered. The returned $\chi^{2}$ values are relatively smaller and closer to the corresponding values of the Y-string model Eq.~\eqref{total}. A consistent physical realization of these observations is that the fluctuations of the node within the static baryon broaden largely enough to give similar effects to a hypothetical $\Delta$-string circumventing the $3Q$ triangle.       

  The Y-string model Eq.~\eqref{total} provides a good description of the lattice numerical data as indicated in Fig.~\ref{p2} and Tables~I-II. Since the model's formula contains a term encoding the effects of the fluctuations of the baryonic node, we could understand by drawing a comparison with the fit behavior of the other ansatzes, Eq.~\eqref{d} and Eq.~\eqref{y}, to what extent the string fluctuations are relevant at the short and long distance scales.

\section{Conclusion}
 The static $3Q$ potential has been studied in lattice QCD at finite temperature using Polyakov loops as a quark source operator. We discussed the potential energy of a Y-string at finite temperature with a fit analysis to the lattice data. The Y-string model provides a suitable description of the numerical data for fit ranges commencing from distances greater than 0.5 fm. The Y-string formula includes a three-body linear confinement term $\sigma_{\rm 3Q} L_{\rm Y}$ in addition to a sub-leading logarithmic term due to the string's Gaussian fluctuations. We found the sub-leading corrections are necessary to reproduce the mesonic string tension measured with the corresponding free string model formula~\cite{PhysRevD.85.077501}.

 At finite temperature, the baryonic gluonic fields are always of a filled $\Delta$-type~\cite{Bakry:2015csa,Bakry:2014gea} with characteristics that have been found to be consistent with the Y-string model describing a system of fluctuating strings~\cite{Bakry:2014gea}. We found here that the data corresponding to the $3Q$ potential match the Y-string model as well. These two results could be of significant interest to the study of QCD strings and the gluonic fields in general and may be promoted into a novel picture for the properties of the confining force in the baryon even at low and zero temperatures. 
\section*{Acknowledgment}

 This work has been funded by the Chinese Academy of Sciences President’s International Fellowship Initiative, Grant No. 2015PM062, NSFC Grants (No. 11035006, No. 11175215, No. 11175220), and the Hundred Talents Program of the Chinese Academy of Sciences (Y101020BR0).
\bibliography{Biblio}
\end{document}